\documentclass[aps,prd,10pt,notitlepage,nofootinbib,superscriptaddress,showkeys,showpacs]{revtex4-1}
\pdfoutput=1
\usepackage{amstext,amsmath,amssymb,amsfonts,bbm,euscript,color}
\usepackage[latin1]{inputenc}
\usepackage{epsfig}
\usepackage{hyperref}
\usepackage{amsthm}
\usepackage{subfigure}
\usepackage{color}
\usepackage{multirow}
\usepackage{tikz}

\usetikzlibrary{arrows,backgrounds}

\usepackage{verbatim}

\usepackage{graphicx}

\usepackage{latexsym}

\newcommand{\bea}{\begin{eqnarray}}	
\newcommand{\eea}{\end{eqnarray}}
\newcommand{\be}{\begin{equation}}	
\newcommand{\ee}{\end{equation}}
\newcommand{\beq}{\begin{equation}}	
\newcommand{\eeq}{\end{equation}}

\newcommand{\Z}{{\mathbb Z}}
\newcommand{\C}{{\mathbb C}}

\def\R{\relax\ifmmode {\mathbb R}  \else${\mathbb R}$\fi}
\def\C{\relax\ifmmode {\mathbb C}  \else${\mathbb C}$\fi}
\def\Z{\relax\ifmmode {\mathbb Z}  \else${\mathbb Z}$\fi}
\def\N{\relax\ifmmode {\mathbb N}  \else${\mathbb N}$\fi}
\def\I{\relax\ifmmode {\mathbb I}  \else${\mathbb I}$\fi}

\begin{document}

\title{Quantum spacetime and the renormalization group: 
Progress and visions}


\author{Antonio D. Pereira}\email{adpjunior@id.uff.br}

\affiliation{ Instituto de F\'isica, Universidade Federal Fluminense, Campus da Praia Vermelha, Av. Litor\^anea s/n, 24210-346, Niter\'oi, RJ, Brazil}

\affiliation{Institute for Theoretical Physics, University of Heidelberg, Philosophenweg 12, 69120 Heidelberg, Germany}


\begin{abstract}
The quest for a consistent theory which describes the quantum microstructure of spacetime seems to require some departure from the paradigms that have been followed in the construction of quantum theories for the other fundamental interactions. In this contribution we briefly review two approaches to quantum gravity, namely, asymptotically safe quantum gravity and tensor models, based on different theoretical assumptions. Nevertheless, the main goal is to find a universal continuum limit for such theories and we explain how coarse-graining techniques should be adapted to each case. Finally, we argue that although seemingly different, such approaches might be just two sides of the same coin. 
\
 

\end{abstract}

\maketitle


\section{Introduction}

The construction of the Standard Model (SM) of particle physics under the perturbative and continuum quantum-field theoretic dogmas led to a successful theory beyond dispute. The detection of the Higgs \cite{Chatrchyan:2012xdj} and the absence of new physics at the LHC so far have crowned the SM as a very accurate quantum description of the fundamental interactions but gravity. However, there are some puzzles which are not addressed by the SM as, e.g., neutrino masses and dark matter. Moreover, the SM is not a fundamental quantum field theory valid up to arbitrarily short scales due to the existence of a Landau pole being thus valid up to some cutoff scale, \cite{GellMann:1954fq,Frohlich:1982tw,Gockeler:1997dn,Gies:2004hy}. 

The classical dynamics of spacetime is successfully described by General Relativity (GR). The recent direct detection of gravitational waves emitted by black holes binary systems \cite{Abbott:2016blz} provides a new arena to test the dynamics of the gravitational field in the strong regime, where its non-linear effects cannot be disregarded\footnote{See also the very recent results from the Event Horizon Telescope in \cite{Akiyama:2019cqa}, in very good agreement with GR predictions.}. So far, GR has shown to be a very good description of it. Given the extraordinary success of those theories, the most natural attitude in order to build a quantum theory of the gravitational interaction is to apply the perturbative quantization techniques of continuum quantum field theory (QFT) to GR. This is the program which is by now referred to as perturbative quantum gravity. As it is known, such a theory is not perturbatively renormalizable meaning that infinitely many terms are needed to cancel ultraviolet (UV) divergences of the theory. Such terms come with arbitrary coefficients which are fixed by experimental data. Therefore, the underlying QFT is not predictive. Nevertheless, the UV divergences can be regularized by the introduction of a UV cutoff and the theory can provide sensible results up to the cutoff scale, since higher-order terms are suppressed by the UV-cutoff. However, at the cutoff scale, such a suppression is lost together with predictivity. For the description of very short distances, a fundamental theory (UV-complete) is necessary.

At this point, many different routes can be taken. In fact, the existence of several different approaches to quantum gravity \cite{Oriti:2009zz} is due to very different underpinning theoretical assumptions. The lack of experimental data as well as clear ``quantum-gravity observables" makes the task of rulling out quantum-gravity models much more subtle. On the other hand, it is even difficult, sometimes, to find a common language for different approaches such that is obviously clear whether the same physical quantity is being computed from different perspectives. 

Irrespective of the chosen approach, a theory of quantum gravity aims at making sense of the following path integral\footnote{Of course, this assertion is restricted to path-integral approaches to quantum gravity.},
\begin{equation}
Z = \sum_{\mathcal{T}}\int \left[\mathcal{D}\mathrm{(geometries)}\right]\mathrm{e}^{i S_{\mathrm{grav}}}\,,  
\label{qgpathint}
\end{equation}
where the functional integral is performed over all geometries, $S_{\mathrm{grav}}$ is the classical or microscopic action encoding the gravitational interaction and the discrete sum is over different topologies. This is just a pictorial representation of the quantum-gravity path integral and should be adapted to each approach accordingly. Moreover, although we do not write explicitly, the path integral might also include matter degrees of freedom and depending on the approach one wants to follow, those are fundamentally important for the consistency of the quantum theory of gravity.

In this contribution, we will review two seemingly different perspectives for the definition of \eqref{qgpathint}. The first one takes the continuum QFT point of view but instead of insisting on the standard perturbative quantization algorithm, looks for a quantum-scale invariant regime driven by an interacting fixed-point in the renormalization group flow. Due to scale invariance, one can zoom in up to arbitrarily short distances without running into divergences. This perspective has witnessed intense progress in the last few years and is known as asymptotically safe quantum gravity (ASQG). The other viewpoint we will review can be seen as a generalization of the matrix model program for two-dimensional quantum gravity to higher dimensions. It goes under the name of tensor models. In this approach, the path integral for quantum gravity is regularized by a lattice-regularization procedure and the integral over geometries is replaced by an integral over tensors which are dual to the building blocks which discretize a given (pseudo-)manifold. The gluing of such building blocks comes with the combinatorially non-trivial interactions of such models. Ultimately, one looks for a continuum limit where the number of building blocks goes to infinity while their volume shrinks to zero and is able to generate a structure which resembles our Universe at large distances.

These approaches take very different routes in their formulation of a consistent description of quantum spacetime. Nevertheless, it is conceivable that despite very different in their construction, they are equivalent, in the sense that they provide the same physics. As an analogy, one can take Yang-Mills theories. In the deep infrared (IR), where the theory becomes strongly coupled, one can use Monte-Carlo simulations to compute correlation functions at the non-perturbative regime. Alternatively, continuum methods as, e.g., functional methods can be employed to reproduce such results. Of course, each method has its own advantages and pitfalls, but being different formulations of the same physics, they can be used conveniently depending on the problem of interest. 

The very construction of each aforementioned approach  depends on the existence of a well-defined continuum limit. In ASQG, the scale-invariant regime allows for a consistent removal of the UV-cutoff introduced to regularize the path integral. In tensor models, the continuum limit is related to the very large number of building blocks used to discretize geometry. Therefore, we look for a mathematical tool which allows us to probe different scales of the theory. Our goal is to find a coarse-graining toolbox which can be applied to quantum gravity in its different formulations. Our choice is the functional renormalization group (FRG), a flexible and powerful framework which allows for a practical implementation of the Wilsonian renormalization program to QFTs. Despite of being formulated very differently, the FRG can be adapted to ASQG as well as to tensor models as we will review later on. Universal quantities, such as critical exponents, can be computed with the FRG and an explicit comparison can be established for ASQG and tensor models. It allows an explicit comparison between the universality classes such theories would fall in. This consists in a very first step towards the establishment of a possible connection between different approaches to quantum gravity. In the case those theories belong to the same universality class, many important issues can be enlightened. As previously explained in the case of Yang-Mills theories, such an equivalence would allow us to use the strengths of each approach to converge towards a consistent picture of the quantum microstructure of spacetime.  
 
This article is organized as follows: In Sect.~\ref{FRGoverview} we provide a concise review about the FRG, fixed points and critical exponents. In Sect.~\ref{ASQGoverview}, a short description about the asymptotic safety scenario for quantum gravity is provided together with a brief summary of some recent progress in the field. In Sect.~\ref{BIRGFoverview}, the tensor models are introduced with a view towards quantum gravity. The main focus is the construction of a practical tool for the discovery of continuum limits for such models. We argue that the FRG can be adapted to this setting and serve as an exploratory tool for continuum limits. In Sect.~\ref{Visions} we summarize our visions regarding the use of renormalization group tool to quantum gravity and advocate that it can be a key element to bridge the gap between different approaches. Consequently, different strengths of different approaches can be combined in a description of the quantum structure of spacetime. 

\section{Functional renormalization group: brief overview} \label{FRGoverview}
\begin{figure}[t]
\begin{center}
\includegraphics[width=0.4\textwidth]{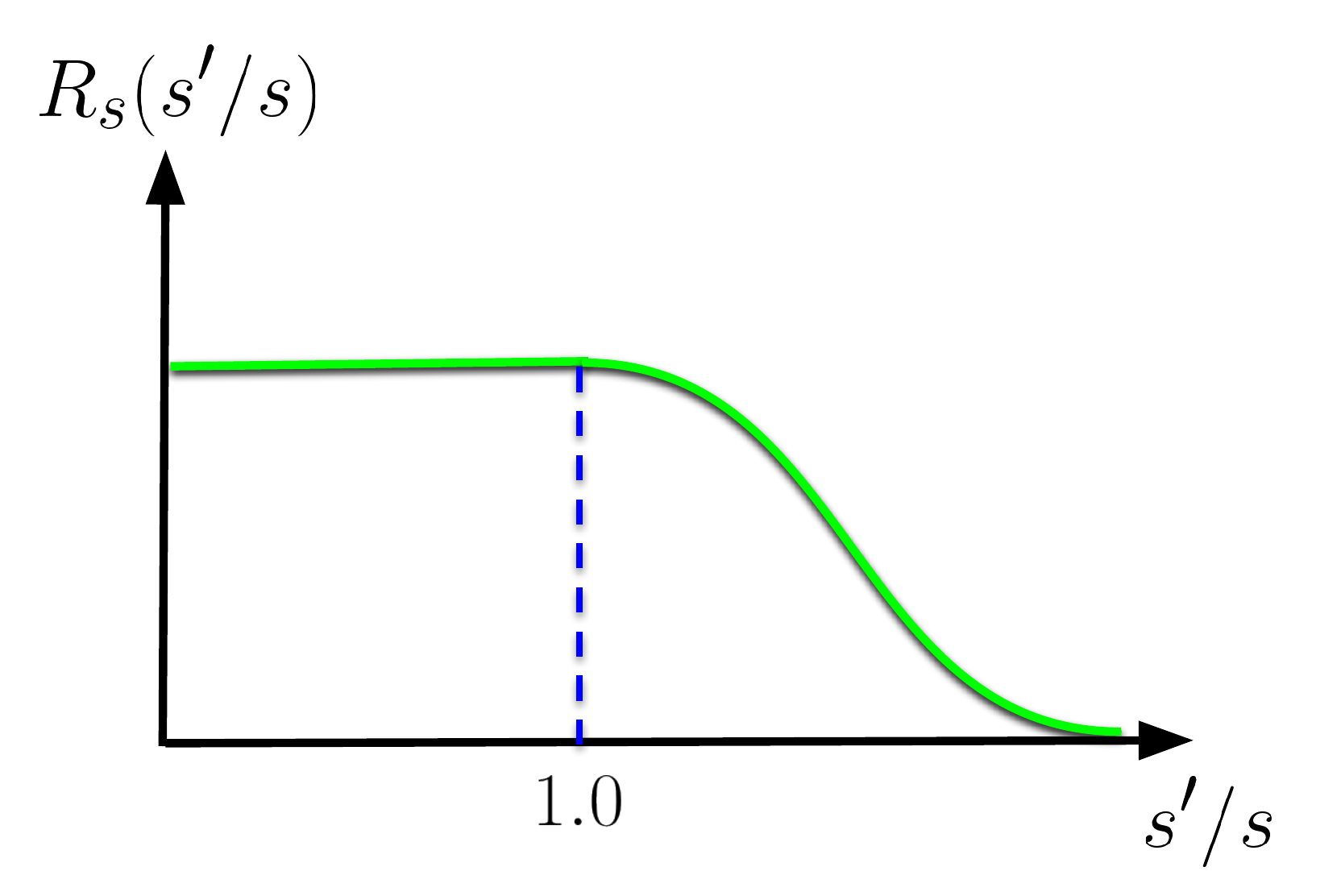}
\caption{Shape of the regulator kernel $R_s$ for a fixed value of $s$.}
\label{regulatorfig}
\end{center}
\end{figure}

Consider a quantum field theory defined by the Euclidean path integral,
\begin{equation}
\mathcal{Z}[J] = \int \left[\mathcal{D}\varphi\right]_{\Lambda_{\mathrm{UV}}}\mathrm{e}^{-S [\varphi]+\varphi \cdot J}\,,
\label{frg1}
\end{equation}
with $\varphi$ being the field content of the theory (not necessarily a simple scalar field), $J$ are the sources coupled to $\varphi$ and $\varphi \cdot J$ denotes the contraction of all indices (spacetime and internal) of the field with the source. The parameter $\Lambda_{\mathrm{UV}}$ corresponds to a UV cutoff\footnote{This is very general and is not necessarily a cutoff in energy. As we will see in the case of tensor models, this parameter is associated to the size of the tensor being thus a dimensionless parameter.} introduced to make the functional measure well-defined. The Wilsonian renormalization perspective incorporates quantum effects by evaluating \eqref{frg1} not at once, but step by step. Using standard QFT on flat spacetime, this can be phrased as the integration of modes with a given momentum shell by shell. The path integral is completely computed by integrating all modes. For each integration step, the result can be expressed in terms of an effective action which takes into account the effects of the integrated modes. Iterating this procedure from $\Lambda_{UV}$ to zero should be completely equivalent to calculate the full path integral at once. The FRG \cite{Morris:1998da,Berges:2000ew,Aoki:2000wm,Pawlowski:2005xe,Gies:2006wv,Delamotte:2007pf,Rosten:2010vm}implements the Wilsonian idea in a smooth way, by modifying the path integral through the introduction of a regulator, as follows,
\begin{equation}
\mathcal{Z}_{s}[J] = \int \left[\mathcal{D}\varphi\right]_{\Lambda_{\mathrm{UV}}}\mathrm{e}^{-S [\varphi]+\varphi \cdot J-\frac{1}{2}\varphi \cdot R_s \cdot \varphi}\,.
\label{frg1a}
\end{equation}
A quadratic term on the fields is introduced with a kernel $R_s$. The field configurations are organized according to the parameter $s$ (at this level, this is completely general) and the kernel $R_s$ implements the suppression of all field configurations which are labeled by a value smaller than $s$. In other words, the quadratic term introduced is such that the path integral is evaluated only over a shell from $\Lambda_{\mathrm{UV}}$ to $s$. In this sense, the parameter $s$ plays the role of a IR cutoff. For $s\to\Lambda_{\mathrm{UV}}$, all configurations are suppressed while for $s\to 0$, all configurations are taken into account corresponding thus to the complete evaluation of the path integral. The smooth suppression can be implemented by any function $R_s$ provided that for any $s^\prime$ such that $s^\prime < s$, $R_s (s^\prime/s) > 0$ and $R_{s}(s^\prime/s) = 0$ for $s^\prime /s$ suffitciently large. Pictorially, the regulator kernel $R_s$ should have the form displayed in Fig.~\ref{regulatorfig} for a fixed value of $s$.
It is very simple to show that the \textit{effective average action} $\Gamma_s [\phi]$ defined by
\begin{equation}
\Gamma_s [\phi] = \mathrm{sup}_{J}\left(\mathrm{Tr}J\cdot \phi - \mathrm{ln}\mathcal{Z}_s\right)-\frac{1}{2}\phi\cdot R_s \cdot \phi\,,
\label{frg2}
\end{equation}
with $\phi = \langle \varphi \rangle_J$, satisfies the exact flow equation,
\begin{equation}
s\partial_s \Gamma_s [\phi] = \frac{1}{2}\mathrm{Tr}\left[\left(\frac{\delta^2 \Gamma_s [\phi]}{\delta \phi^2}+ R_s\right)^{-1}s\partial_s R_s \right]\,.
\label{frg3}
\end{equation}
By definition, $\Gamma_{s=\Lambda_{\mathrm{UV}}} = S$ and $\Gamma_{s=0} = \Gamma$, with $\Gamma$ being the full effective action of the underlying QFT. Hence, the effective average action $\Gamma_s$ interpolates between the classical/microscopic action $S$ and the effective action $\Gamma$ which takes all the quantum fluctuations into account. Eq.(\ref{frg3}) is known as the FRG equation, flow equation or Wetterich equation. It is exact and has the a very simple structure, being thus very useful for practical purposes. In fact, it has the same structural form of the one-loop contribution to the effective action in perturbation theory, a fact which is exactly preserved due to the quadratic form of the regulator.  

The flow equation can be seen as an alternative to the path integral: Instead of performing the functional integral in order to assess the quantum action which generates all the building blocks of a given QFT, one can solve the flow equation with the initial condition being given by the classical action. Therefore, in general, solving the flow equation exactly is as difficult as evaluating the path integral completely and approximations are needed. Nevertheless, let us assume that we are able to compute the effective average action. It is expanded on a basis of (quasi-)local operators $\mathcal{O}(\phi)$ of the fields $\phi$ as
\begin{equation}
\Gamma_s [\phi] = \sum_i \bar{g}_{i} (s) \mathcal{O}^i (\phi)\,,
\label{frg4}
\end{equation}
with $\bar{g}_{i} (s)$ being called the running coupling constant associated to $\mathcal{O}^{i}(\phi)$. Typically, this basis is infinite-dimensional\footnote{When integrating the modes step by step in the path integral, one realizes that all terms compatible with the symmetries of the theory are generated in the effective action - apart from anomalies which we ignore in this article.}. Plugging \eqref{frg4} in the flow equation \eqref{frg3} yields
\begin{equation}
s\partial_s\Gamma_s [\phi] = \sum_i (s\partial_s \bar{g}_i (s)) \mathcal{O}^{i} (\phi)\,.
\label{frg5}
\end{equation}
Expanding the righthand side of eq.\eqref{frg3} on the same basis $\left\{\mathcal{O}^i (\phi)\right\}$, one can read off the coefficients $s\partial_s g_i (s) = \bar{\beta}_{g^i}$. Such coefficients are known as beta functions of the \textit{dimensionful} couplings. This leads to a system of infinitely many coupled equations which, typically, is not autonomous. The system can be, sometimes, made autonomous by rescaling all the couplings with an appropriated power of $s$, i.e.,
\begin{equation}
\bar{g}_i (s) = s^{\left[\bar{g}_i\right]}g_i (s)\,.
\label{frg6}
\end{equation}
The number $\left[\bar{g}^i \right]$ is known as the \textit{canonical dimension} of the coupling $\bar{g}^i$ and $g^i (s)$ are called \textit{dimensionless} couplings. They parametrize an infinite-dimensional space called \textit{theory space}. In standard QFTs formulated on a background spacetime, the canonical dimension is immediately extracted by dimensional analysis (mass dimension associated to each couplings). However, as explained before, the parameter $s$ might be more abstract than a momentum scale and even  dimensionless. In such a case, we fix the canonical dimensions by demanding that the resulting system of equations obtained from the flow equation is autonomous. Thence, the beta functions of the dimensionful couplings are expressed as
\begin{equation}
\bar{\beta}_{g^i} = \left[\bar{g}^i \right]s^{\left[\bar{g}_i\right]}g_i (s)+s^{\left[\bar{g}_i\right]}\beta_{g^i}\,,
\label{frg7}
\end{equation}
with $\beta_{g^i} = s\partial_s g^i$ being the beta functions of the \textit{dimensionless} couplings (which we will simply refer to as beta functions). The first term on the righthand side of \eqref{frg7} is associated to the canonical dimension of the corresponding coupling while the second term arises due to the non-trivial dependence on the parameter $s$, i.e., due to quantum effects. Ultimately, we are interested in a UV completion for quantum gravity. A UV complete theory is such that all dimensionless couplings are finite at $s\to\Lambda_{\mathrm{UV}}\to\infty$. A sufficient condition for finiteness is that the system of beta functions  admits a fixed point, i.e., a point in theory space $g^\ast = \left\{g^{\ast}_{1},g^{\ast}_{2},g^{\ast}_{3},\ldots\right\}$ (with finite values for all couplings) where all beta functions vanish, $\beta_{g^i}(g^\ast) = 0\,,\forall i$. At the fixed point, the running of the couplings with $s$ ceases and the theory reaches a \textit{scale-invariant regime} \cite{Wetterich:2019qzx}. At this point, the limit $s\to\Lambda_{\mathrm{UV}}\to\infty$ can be safely taken.

\begin{figure}[t]
\begin{center}
\includegraphics[width=0.4\textwidth]{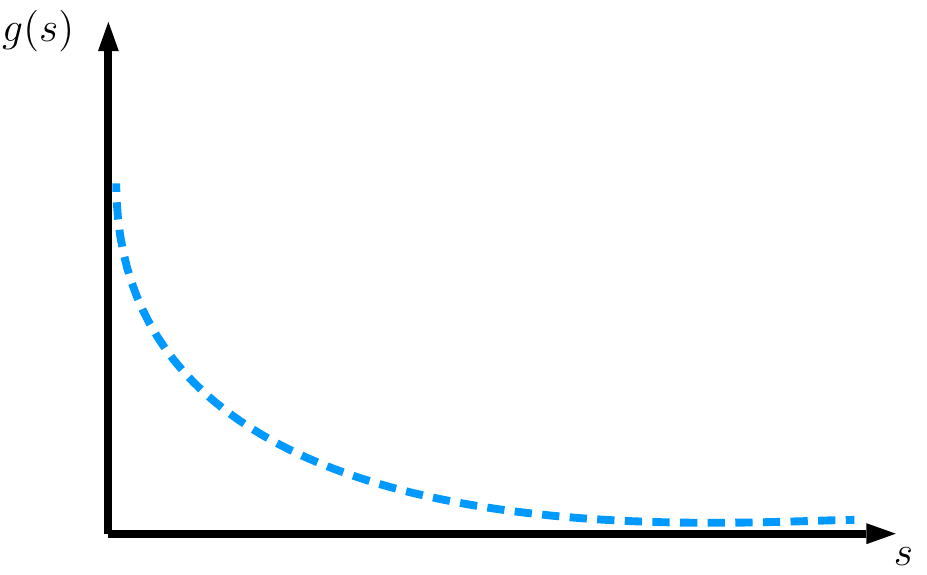}\,\,\,\,\,\,\,\,\,\,\,\,\,\,\,\,
\includegraphics[width=0.4\textwidth]{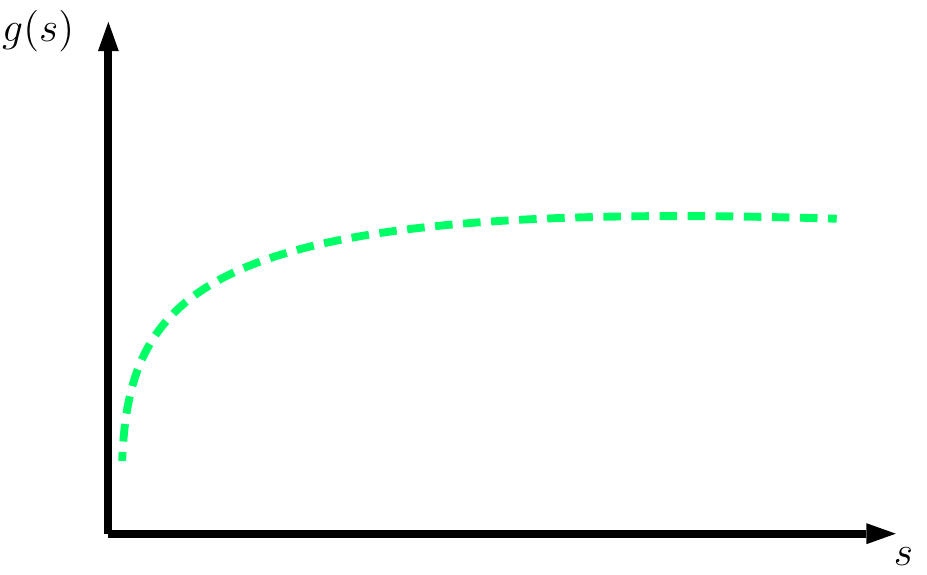}
\caption{On the left plot, a representation of an asymptotically free coupling. In this case, the fixed point occurs at vanishing coupling. The plot on the right shows an asymptotically safe coupling which attains a non-trivial but finite value in the UV.}
\label{couplingsfig}
\end{center}
\end{figure}

A particularly well-known case of fixed point is the asymptotically free one. In this case, for arbitrarily large values of energy, the couplings tend to zero, i.e., they reach vanishing value at the fixed point. However, one could think of a fixed point at non-vanishing (non-trivial) values\footnote{Oftenly, the asymptotically free fixed point is called Gaussian or non-interacting fixed point and the asymptotically safe one, non-Gaussian or interacting fixed point.}. This would correspond to a well-defined UV completion of the theory. In this case, we refer to the theory as \textit{asymptotically safe}. We refer to Fig.~\ref{couplingsfig} for a pictorial representation of asymptotically free and safe couplings.

While asymptotically free fixed points are well described in perturbation theory, non-trivial fixed-points can be impossible to be accessed within this framework. In fact, if the location of the non-Gaussian fixed point is too far from the origin of the theory space, then the couplings can assume large values, which invalidates the perturbative assumption. As such, non-perturbative techniques are required. Technically, this is the reason behind the difficulty to probe non-Gaussian fixed points, in general. Fortunately, although the FRG requires an approximation to be useful at the practical level, different schemes which go beyond the perturbative expansion are possible. Hence, even within approximations, the FRG can be a valuable tool to probe interacting fixed points. Among the different approximation methods to the FRG, the one we will be mostly interested in this article consists in truncating the infinite-dimensional theory space to a subspace (it does not need to be finite dimensional necessarily). Within this subspace, an ansatz for the effective average action is proposed and the righthand side of the flow equation can be evaluated. The result will typically leak from the original subspace and a projection rule must be defined. Thus, the flow is projected to the truncated subspace and one can immediately read off the beta functions. If no external input is given, the quality of the truncation is tested by enlarging the subspace and testing the stability of the results. The more convergent the results are for arbitrary enlargements of the truncation, the better the truncation is. As can be guessed, if no guiding principle is discovered, setting a ``good" truncation is not a simple task and requires several consistency tests.

\begin{figure}[t]
\begin{center}
\includegraphics[width=0.5\textwidth]{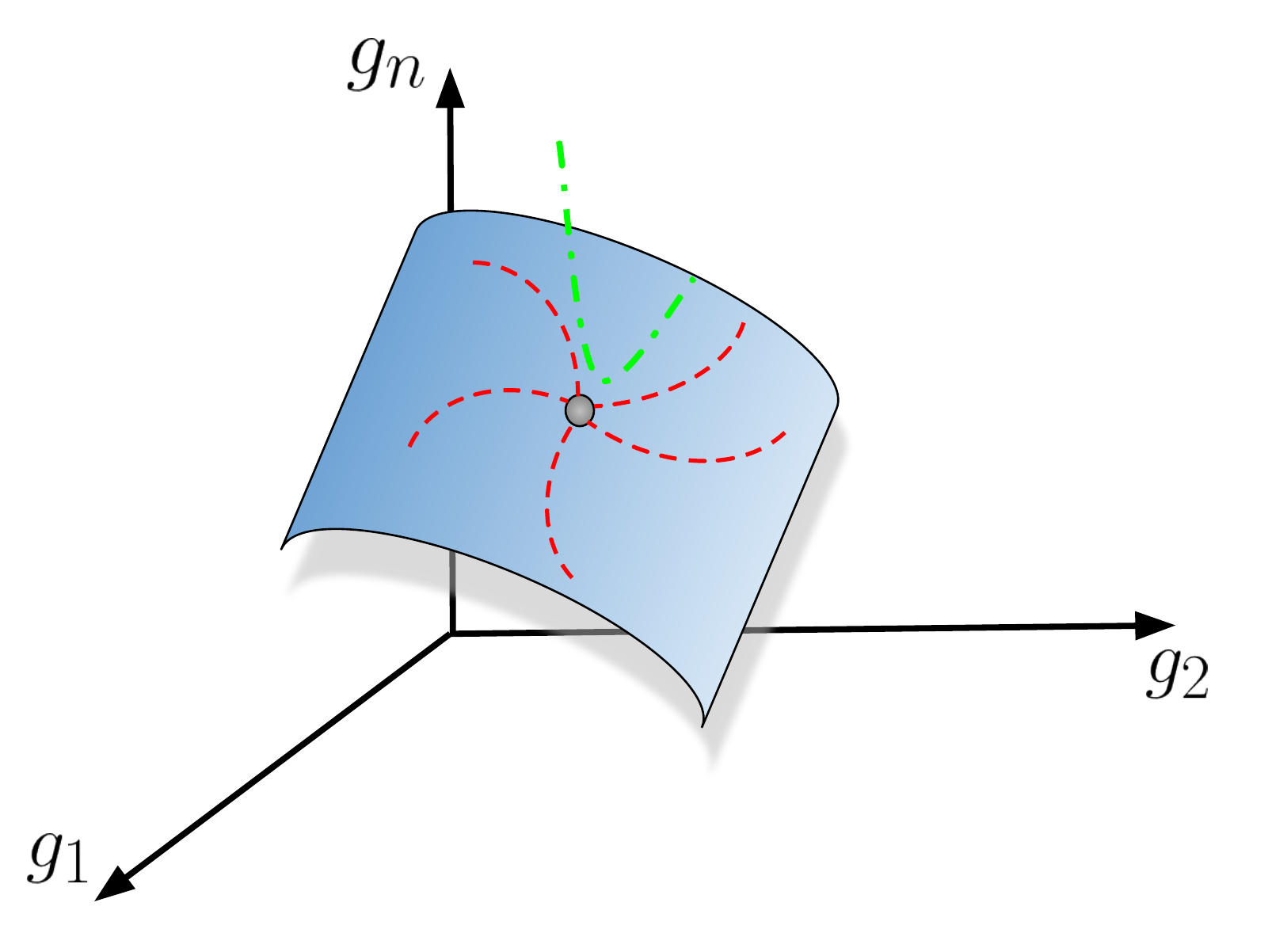}
\caption{The surface represents the critical surface associated to the fixed point. Trajectories which hit the fixed point (the red ones ) span the suface - defined by the UV attractive/relevant directions. The green trajectory is an example of a deviation to a UV-repulsive direction.}
\label{theoryspacefig}
\end{center}
\end{figure}

Let us assume that we managed to find an interacting fixed point for a given QFT. A natural question which arises is: Since the theory space is, in general, infinite dimensional, do we actually need to provide infinitely many boundary conditions to set up a renormalization group trajectory which hits the fixed point? The answer is: typically, no. In order to argue that, let us consider the fixed point $g^\ast = \left\{g^{\ast}_1,g^{\ast}_2,\ldots\right\}$ and the linearized flow around it, i.e., 
\begin{equation}
s\partial_s (g_i - g^{\ast}_i) = \sum_j \frac{\partial \beta_{i}}{\partial g_j}\Bigg|_{g = g^{\ast}} (g_ j - g^\ast_j)\equiv \sum_j \mathcal{M}_{ij}(g_ j - g^\ast_j)\,, 
\label{frg8}
\end{equation}
where $\mathcal{M}_{ij}$ is known as stability matrix. It can be diagonalized by a suitable change of coordinates $z_i = \sum_j S_{ij}(g_j - g^{\ast}_j)$ leading to
\begin{equation}
s\partial_s z_i = \lambda_i z_i\,,
\label{frg9}
\end{equation}
with $\lambda_i$ being the eigenvalues of $\mathcal{M}_{ij}$. It is convenient to define the renormalization group ``time" $t = \mathrm{ln}~s/s_0$, with $s_0$ being a reference value. Then, eq.\eqref{frg9} can be solved,
\begin{equation}
z_i (t) = C_i\, \mathrm{e}^{\lambda_i t}\,,
\label{frg10}
\end{equation}
where $C_i$ are integration constants. Changing back to the original coordinates, eq.\eqref{frg10} is expressed as 
\begin{equation}
g_i (s) = g^\ast_i +\sum_j V_{ij}\mathrm{e}^{\lambda_j t}\,,
\label{frg11}
\end{equation}
with $V_{ij}$ being constants. Defining the so-called \textit{critical exponents} $\theta_i = -\lambda_i$, eq.\eqref{frg11} becomes
\begin{equation}
g_i (s) = g^\ast_i +\sum_j V_{ij}\left(\frac{s}{s_0}\right)^{-\theta_j}\,.
\label{frg12}
\end{equation}
From eq.\eqref{frg12}, one sees that if $\theta_j >0$, then the flow approaches to the fixed point in the UV ($s\to\Lambda_{\mathrm{UV}}\to \infty$) irrespective of the value of $V_{ij}$. Hence, for those directions the $V_{ij}$ are free parameters. Towards the infrared (decreasing values of $s$), the distance between the coupling $g_i$ and $g^{\ast}_i$ grows. Hence, we say that $\theta_j > 0$ defines a UV attractive/IR repulsive direction. Such positive critical exponents define the so-called \textit{relevant directions}. On the contrary, if $\theta_j < 0$, the flow of the corresponding coupling is driven away from the fixed-point value. In this case, in order to hit the fixed point, the coefficient $V_{ij}$ must vanish. Such directions are the so-called \textit{irrelevant directions} and are UV repulsive/IR attractive. If the critical exponent vanishes, then the direction is called \textit{marginal}. In this case, one needs to go beyond the linearized order to check whether the direction is marginally relevant, irrelevant or exactly marginal. Towards the IR, relevant directions are associated with free parameters while irrelevant ones do not. Therefore, we should fix those free parameters with experimental data. Consequentely, if the number of relevant directions is finite, then a finite number of experiments should be performed to fix all free parameters. This ensures that the theory is predictive. The relevant directions define a hypersurface in theory space known as \textit{critical surface}. Predicitivity is achieved if the hypersurface is finite dimensional. We refer to Fig.~\ref{theoryspacefig} for a cartoon representing the hypersurface in theory space. 

If the fixed point is trivial (Gaussian), then the number of relevant directions is dictated by the number of couplings with positive and zero (depending whether it is marginally relevant or not) canonical dimension in usual QFTs. In the language of perturbation theory, the critical surface is spanned by the power-counting renormalizable couplings. However, for a non-Gaussian fixed point, the situation is more complicated. In fact, canonically relevant directions can turn to irrelevant ones at the fixed point and vice-versa due to the interacting nature of the fixed point. If the fixed-point is near-Gaussian, i.e., if its location is sufficiently close to the Gaussian fixed point, then it is expected that the canonical dimension works as a reasonably good guidance for the dimensionality of the critical surface. In this case, perturbation theory should be applicable despite of the interacting nature of the fixed point. In summary, the existence of a fixed-point of the renormalization group flow which features a finite number of relevant directions prevents the couplings of the theory to diverge and ensures predictivity. Since this property might be probed just beyond perturbation theory, one might want to call such a theory non-perturbatively renormalizable. We will simply call such a theory as an asymptotically safe theory.

\section{The asymptotic safety scenario for quantum gravity} \label{ASQGoverview}

In this section, we present a very brief overview about the asymptotic safety paradigm in the context of quantum gravity \cite{Percacci:2011fr,Reuter:2012id,Eichhorn:2017egq,Percacci:2017fkn,Eichhorn:2018yfc,Reuter:2019byg}. As is well known, the perturbative quantization of GR as a QFT for the metric leads to a (perturbatively) non-renormalizable theory. In this case, the path integral is formally expressed as\footnote{We restrisct ourselves to the Euclidean path integral. The choice is for technical reasons. We shoudl emphasize that establishing whether what is going to be presented remains valid in the Lorentzian setting is still a challenging open question.}
\begin{equation}
\mathcal{Z}=\int \left[\mathcal{D}g_{\mu\nu}\right]\mathrm{e}^{-S_\mathrm{EH}}\,,
\label{asqg1}
\end{equation}
with $S_{\mathrm{EH}}$ being the Einstein-Hilbert action, given by
\begin{equation}
S_{\mathrm{EH}} = \frac{1}{16\pi G_N}\int \mathrm{d}^4x \sqrt{g}\left(2\Lambda - R\right)\,, 
\label{asqg2}
\end{equation}
where $\Lambda$ and $G_N$ denote the cosmological and Newton constants. The quantization usually employs the background field method \cite{Percacci:2017fkn}, where a fiducial metric $\bar{g}_{\mu\nu}$ is introduced. Then, the complete fluctuating metric is split as
\begin{equation}
g_{\mu\nu} = \bar{g}_{\mu\nu} + h_{\mu\nu}\,,
\label{asqg3}
\end{equation}
where $h_{\mu\nu}$ stands for the quantum fluctuations about the fixed metric $\bar{g}_{\mu\nu}$. This is usually called ``linear split" of the metric. Although the separation between background and fluctuation is linear, such a choice has some drawbacks as, for instance, it does not preserve the signature of the metric if $h_{\mu\nu}$ is allowed to fluctuate widely.  Different choices of parametrization as, e.g., the exponential parametrization, circumvent some of those issues. See, e.g. \cite{Kalmykov:1995fd,Kalmykov:1998cv,Nink:2014yya,Gies:2015tca,Ohta:2016npm,Ohta:2016jvw,Goncalves:2017jxq,Ohta:2018sze,deBrito:2018jxt} for discussions on different choices of parametrization of the quantum fluctuations in perturbative quantum gravity as well as in asymptotic safety. However, typically, this comes with a non-linear separation between background and fluctuation. In perturbation theory, the fluctuation $h_{\mu\nu}$ is taken to be small. Being an integral over metrics, we need to ``gauge fix" it in order to perform a functional integral over \textit{geometries}. This can be achieved by the usual Faddeev-Popov procedure. This entails the introduction of a gauge-fixing term together with a Faddeev-Popov ghosts term. It is not the scope of this article to explain such things in details. For a recent pedagogical introduction to that, we refer to \cite{Percacci:2017fkn}. Introducing all that, we can perform a standard calculation of the divergences order by order in perturbation theory. In \cite{tHooft:1974toh,Christensen:1979iy}, the one-loop divergences were computed for the Einstein-Hilbert action without and with cosmological constant. The logarithmic divergence contains terms that are not part of the microscopic action and they are proportional to background curvature squared terms. Nevertheless, if matter is absent, those terms can be dealt with on-shell. If we disregard the cosmological constant, then the theory is Ricci flat and the one-loop divergence vanishes on-shell. If a non-vanishing cosmological constant is included, then the Ricci tensor is related to the cosmological constant and the one-loop divergence can be absorbed by a redefinition of $\Lambda$. This means that the theory, i.e., quantum GR is one-loop renormalizable if matter is absent.  In the presence of matter, however, the on-shell condition produces terms which are proportional to the energy-momentum tensor. Such a structure is absent in the microscopic action and, therefore, the theory is not renormalizable at one-loop. If one insists on pure gravity, the natural next step is to check whether divergences will pop up at two-loops. In \cite{Goroff:1985th}, it was shown that a non-vanishing divergence, proportional to ${\bar{R}^{\mu\nu}}_{\lambda\rho}{\bar{R}^{\lambda\rho}}_{\alpha\beta}{\bar{R}^{\alpha\beta}_{\mu\nu}}$ exists. This term is not present in the bare action and a counterterm with free coefficient should be added in order to absorb this divergence. At every order, in perturbation theory, new counterterms with arbitrary coefficients will be needed to control divergences and this spoils completely the predictivity power of the theory. In fact, this was not an unexpected result due to power-counting arguments, but it was logically possible that such terms could arise with vanishing coefficient. Explicit calculations show that this is not the case.

This fact does not imply that there is a fundamental incompatibility between GR and QFT as is oftenly spread. In fact, one can treat quantum GR as an effective field theory, meaning that it is valid up to some ultraviolet cutoff. Calculations performed below such a scale are perfectly valid. At the cutoff scale, the theory breakdown and begs for a UV completion. Nevertheless, quantum gravitational corrections within such a framework are possible to be computed and the most iconic result consists on the evaluation of quantum corrections to the Newtonian potential, see \cite{Donoghue:1993eb}. The perturbative non-renormalizability of GR can be interpreted in several different ways. One possibility is to assume that the classical action, the Einstein-Hilbert action, should be replaced by another action which renders a perturbatively renormalizable QFT. A celebrated example is higher-derivative gravity introduced in \cite{Stelle:1976gc}. On top of the Einstein-Hilbert terms, $R_{\mu\nu}R^{\mu\nu}$ and $R^2$ contributions are added to the classical Lagrangian. The underlying quantum theory is perturbatively renormalizable and, under certain circumstances, asymptotically free. This would characterize a UV-complete theory amenable to perturbation theory. Nonetheless, this theory contains a ghost state in its spectrum, a fact that might jeopardize unitarity. In fac, this issue is subject of a long-standing debate and different strategies to circumvent such a problem have been explored. We refer to \cite{Anselmi:2017ygm,Anselmi:2018ibi,Anselmi:2018tmf,Anselmi:2018cau} for a very recent account on the unitarity issue in higher-derivative gravity with curvature squared terms (and to \cite{Salvio:2018crh} for a recent review on curvature-squared higher-derivative gravity) and \cite{Modesto:2015ozb} for models with higher-order curvature terms. Thence, the quest for a UV complete theory of quantum gravity within the standard perturbative QFT toolbox is still an active research field. 

Another possibility is to verify whether quantum gravity features a non-trivial ultraviolet fixed point along the renormalization group flow. Following the discussion we have made in the previous section, this would entail a ``non-perturbatively renormalizable" QFT which is free of divergences and predictive, provided that the number of relevant directions at the fixed point remains finite. Hence, quantum gravity would be asymptotically safe. Such a possibility was pointed out and elaborated, for the first time, by Weinberg in \cite{Hawking:1979ig}. However, it was in \cite{Reuter:1996cp}, where the FRG equation was adapted to quantum gravity, that a toolbox was developed for practical verifications of this scenario. The background metric $\bar{g}_{\mu\nu}$ defines a momentum scale by the eigenvalues of the background Laplacian and the field configurations can be organized in momentum shells. This sets up a local coarse-graining procedure and the step by step functional integral is performed by the introduction of a regulator quadratic in the fluctuation fields $h_{\mu\nu}$ as discussed in the previous sections, i.e.,
\begin{equation}
\Delta S_k = \int \mathrm{d}^4x\sqrt{\bar{g}}\,h_{\mu\nu}R^{\mu\nu\alpha\beta}_k (\bar{\nabla}^2)h_{\alpha\beta}\,.
\label{asqg4}
\end{equation}
The regulator kernel $R^{\mu\nu\alpha\beta}_k (\bar{\nabla}^2)$ (not to be confused with curvature) ensures that only modes with momentum (squared) larger than $k^2$ are integrated out. The quadratic structure of \eqref{asqg4} ensures that the flow equation will have the simple one-loop form as in eq.\eqref{frg3}. In \cite{Reuter:1996cp}, a simple choice for the effective average action $\Gamma_k$ was made. It is known as the Einstein-Hilbert truncation and consists in assuming that the effective average action is written as
\begin{equation}
\Gamma^{\mathrm{EH}}_k [g_{\mu\nu},\bar{g}_{\mu\nu}] = \frac{1}{16\pi G_k}\int\mathrm{d}^4x\,\sqrt{g}(2\Lambda_k-R)+S_{\mathrm{gf}}+S_{\mathrm{gh}}\,,
\label{asqg5}
\end{equation}
where $S_{\mathrm{gf}}$ and $S_{\mathrm{gh}}$ are suitable gauge-fixing and Faddeev-Popov terms. The parameters $G_k$ and $\Lambda_k$ are the running and dimensionful Newton and cosmological constants. Another feature regarding eq.\eqref{asqg5} is that the effective average action is written as a functional of the full\footnote{This notation is not accurate. In fact, the metric $g_{\mu\nu}$ that enters as the argument of the effective average action corresponds to the expectation value of the metric $g_{\mu\nu}$ that appears in definition of the path integral \eqref{asqg1}. We employ the same name for both for simplicity.} $g_{\mu\nu}$ and background metrics $\bar{g}_{\mu\nu}$ independently. This is due to the introduction of a gauge-fixing term and the regulator \eqref{asqg4}. Those terms treat the full metric and the fluctuation $h_{\mu\nu}$ in such a way that they do not appear just as $g=\bar{g}+h$. Consequently, the renormalization group flow will generate further terms which do not respect the linear split of the metric and, therefore, it should be projected to a symmetric subspace by demanding that the effective average action satisfies suitable Ward identities. This is way beyond the scope of the present discussion and refer to, e.g., \cite{Eichhorn:2018yfc} and references therein for further details. The conceptual ingredient we wanted to introduce with this discussion is that although the introduction of an auxiliary background $\bar{g}_{\mu\nu}$ has brought up the possibility to define a momentum scale an thus a local coarse-graining procedure, it came with the cost of breaking/deforming the split symmetry between background and fluctuation fields. Thence, background independence, a fundamental requirement in a theory of quantum gravity, is not manifest and should be explicitly verified.  

Remarkably, in the pioneering work \cite{Reuter:1996cp}, a non-trivial fixed point, the Reuter fixed point, in the bidimensional truncated theory space parametrized by the dimensionless Newton and cosmological constants was found. It features two-relevant directions. Concretely, see, e.g., \cite{Codello:2008vh}, the beta functions for the dimensionless Newton and cosmological constants, $g_k \equiv k^2 G_k$ and $\lambda_k = k^{-2} \Lambda_k$ are, respectively, 
\begin{eqnarray}
\beta_g &=& 2g_k - \frac{g^2_k}{3\pi}\frac{11-18\lambda_k+28\lambda^2_k}{(1-2\lambda_k)^2-\frac{1+10\lambda_k}{12\pi}g_k}\,,\nonumber\\
\beta_\lambda &=& -2\lambda_k + \frac{g_k}{6\pi}\frac{3-4\lambda_k-12\lambda^2_k-56\lambda^3_k+\frac{107-20\lambda_k}{12\pi}g_k}{(1-2\lambda_k)^2-\frac{1+10\lambda_k}{12\pi}g_k}\,.
\label{asqg6}
\end{eqnarray}
The fixed-point value is around $\lambda^\ast = 0.2$ and $g^\ast=0.7$ and the critical exponents form a complex conjugate pair with real part of order one. The existence of such a fixed point was robustly verified in much more sophisticated truncations, where higher order curvature terms were took into account. The inclusion of more terms can be done in several different directions in theory space. One example is to consider higher power of the Ricci scalar $R$ or non-polynomial functions of $R$. Explicit calculations with polynomials up to $R^{70}$, see \cite{Falls:2018ylp} show not only that the existence of the fixed point is stable but also it is quantitatively apparently convergent, i.e., its value does not change significantly under improvements of the truncations. Furthermore, the near-canonical scaling probed by the critical exponents suggest that such a fixed point is near-perturbative, see \cite{Eichhorn:2018ydy}. Extensions of the truncations to different directions in theory space were also performed, namely, the inclusion of $R_{\mu\nu}R^{\mu\nu}$ terms and, notably, the Goroff-Sagnotti term. For an incomplete list of works we refer to \cite{Reuter:2001ag,Litim:2003vp,Codello:2006in,Machado:2007ea,Benedetti:2009rx,Benedetti:2009gn,Manrique:2010am,Manrique:2011jc,Christiansen:2012rx,Falls:2013bv,Benedetti:2013jk,Codello:2013fpa,Falls:2014tra,Christiansen:2014raa,Christiansen:2015rva,Gies:2015tca,Gies:2016con,Biemans:2016rvp,Christiansen:2016sjn,Denz:2016qks,Knorr:2017fus,Knorr:2017mhu,Christiansen:2017bsy,Falls:2017lst,Falls:2018ylp}. Remarkably, the number of relevant directions seems to saturate at around three a fact that ensures that the theory is predictive. Of course, a formal proof of the existence of such a fixed point is desirable, but this seems to be an extremely challenging problem. The more realistic strategy is to find evidence for the existence of the fixed point and, so far, all results points towards that. Very schematically, the mechanism which drives asymptotic safety is the compensation between the canonical dimension term in the beta functions (first terms on the righthand side of both equations in \eqref{asqg6}) and the non-trivial quantum corrections. Such corrections are typically non-vanishing away from the free-theory fixed point.

Although pure gravity has shown several different evidence for the existence of a suitable non-trivial fixed point in the renormalization group flow, there is the possibility that matter degrees of freedom, when taken into account, induce contributions to the gravitational beta functions and destroy the existence of the fixed point. Hence, a crucial question to be answered in the asymptotic safety program is whether observed matter degrees of freedom are compatible with the fixed point structure. Fortunately, since the asymptotic safety scenario is described in terms of usual continuum QFT language, coupling matter fluctuations is straightforward. In the recent years, many different works explored that and revealed that the fixed point structure is compatible with the SM matter content. Such an achievement opens the door for the formulation of a consistent description of quantum gravity as a QFT which is compatible with our knowledge about the matter content of our Universe. See \cite{Percacci:2002ie,Percacci:2003jz,Narain:2009fy,Zanusso:2009bs,Eichhorn:2011pc,Eichhorn:2012va,Dona:2013qba,Dona:2014pla,Labus:2015ska,Oda:2015sma,Meibohm:2015twa,Dona:2015tnf,Meibohm:2016mkp,Eichhorn:2016esv,Eichhorn:2016vvy,Biemans:2017zca,Hamada:2017rvn,Christiansen:2017qca,Eichhorn:2017eht,Eichhorn:2017egq,Eichhorn:2017sok,Christiansen:2017cxa,Eichhorn:2017als,Alkofer:2018fxj,Eichhorn:2018akn,Eichhorn:2018ydy,Eichhorn:2018yfc,Eichhorn:2018nda,Pawlowski:2018ixd} for some works on that. Conversely, the impact of quantum-gravity fluctuations to the matter sector can be also analyzed. In fact, the assumption of the existence of an asymptotically safe fixed point in the gravitational couplings leads to several interesting effects in the matter couplings and, in particular, can provide a prediction of the Higgs and top quark masses \cite{Shaposhnikov:2009pv,Pawlowski:2018ixd,Eichhorn:2017muy,Eichhorn:2018whv}, resolve the Landau pole in the SM \cite{Harst:2011zx,Christiansen:2017gtg,Eichhorn:2017lry} and explain the gauge hierarchy problem, see \cite{Wetterich:2016uxm}.

Of course, all this consists a promising picture towards the construction of a quantum theory of gravity (and matter). However, the challenging task of proving the existence of the fixed point as well as how all this fits to a Lorentzian setting and the fate of background independence are important open questions that need to be addressed in this approach. Besides the important recent work done within the continuum formulation of quantum gravity, we advocate in this article that a potential very fruitful route is to use different methods such as lattice simulations or different formulations to probe such properties. The interplay between different frameworks by combining different strengths of each of them could provide new key insights and avoid technical complications that can hamper a convicing check of the aforementioned properties. In the next section, we briefly explain how to define the path integral for quantum gravity using tensor models a how this could be, eventually, connected to asymptotic safety. 

\section{Background-independent renormalization group flows} \label{BIRGFoverview}

Since a formal mathematical proof of the existence of suitable fixed point in the renormalization group flow is beyond our current capabilities, it is extremely desirable (and necessary) to find evidence for its existence from different perspectives. Furthermore, as we mentioned, there are challenges that might be easier to tackle employing a different framework. A concrete example is background independence. In the QFT for metrics, an explicit background is chosen for concrete calculations and the statement of background independence should be phrased as the fact that no special role is played by the choice of background. Nevertheless, the explicit check of that is far from trivial and the issue whether the resulting quantum theory is really background independent is open. On the other hand, if the path integral over geometries could be evaluated with no reference to a background, then a comparison between the resulting quantum theories would be possible and, if they are physically equivalent, background independence would be established for the asymptotic safety scenario. 

\begin{figure}[t]
\begin{center}
\includegraphics[width=0.4\textwidth]{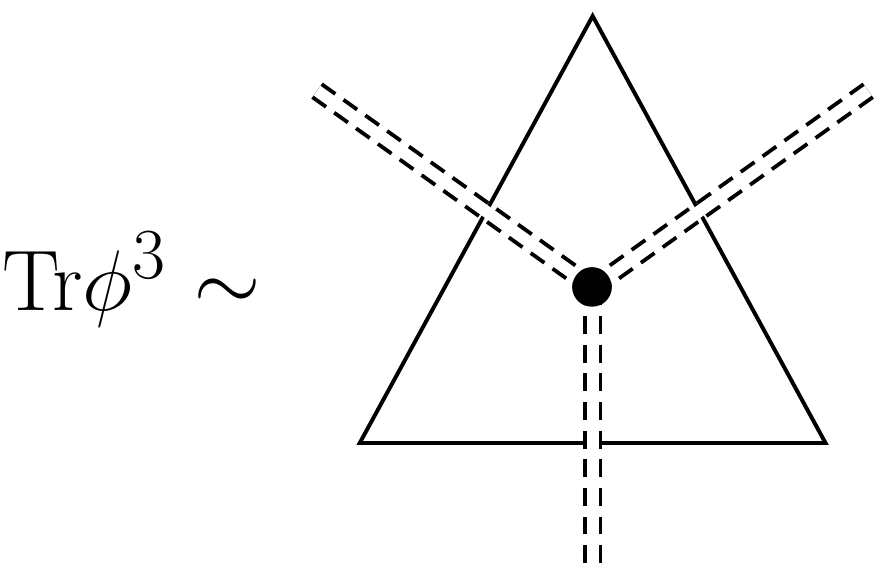}
\caption{Each dashed line is associated to a matrix. They are dual to the edges of a triangle. Feynman diagrams correspond to gluing such vertices with propagators. Geometrically, this corresponds to glue the triangles along their edges.}
\label{dualmatrixfig}
\end{center}
\end{figure}

One possibility is to discretize the path integral in a suitable way, providing a lattice-like regularization, and perform the sum over discretized geometries. The physical content of the theory would then be available in the continuum limit where the lattice spacing shrinks to zero. Of course, it is not clear whether the continuum limit would be well-behaved giving rise to extended four-dimensional geometries at large distances. This program has a sucessful story in two dimensions. There, the sum over geometries and topologies is enconded in matrix models \cite{DiFrancesco:1993cyw}. They correspond to statistical models of random matrices $\phi_{ab}$ of size $N^\prime$ whose interactions are dual to building blocks of geometry. For instance, if one considers a cubic matrix model defined by the partition function
\begin{equation}
\mathcal{Z} \sim \int \left[d\phi\right]\mathrm{e}^{-\frac{1}{2}\mathrm{Tr}\phi^2+\frac{g}{\sqrt{N^\prime}}\mathrm{Tr}\phi^3}\,,
\label{birgf1}
\end{equation}
the interaction is dual to a triangle as is shown in Fig.~\ref{dualmatrixfig}.
The continuum limit should not depend on the choice of the building blocks, i.e., we could have chosen squares intead of triangles as the building blocks, for example. Hence, we aim at finding a universal continuum limit. As is intuitive to understand, the Feynman diagrams of this theory correspond to gluing vertices and, consequently, triangles (or different building blocks). Therefore, the Feynman expansion correspond to a sum of triangulated surfaces. It can be organized by the topology of the resulting manifold, the so-called $1/N^\prime$-expansion, and due to tools particular to two dimensions, this can be completely evaluated. In this case, in order to have contributions from all different topologies, a continuum limit is obtained by tunning the coupling $g$ to a critical value while the matrix model size $N^\prime$ is taken to infinity. This is the so-called double-scaling limit \cite{Douglas:1989ve,Brezin:1990rb,Gross:1989vs,Gross:1989aw}. It is defined by
\begin{equation}
N^\prime (g-g_{\ast})^{5/4} = \mathrm{const}
\label{birgf2}
\end{equation}
with $N^\prime \to \infty$ and $g\to g_\ast$. Intuitively, $N^\prime$ counts the number of degrees of freedom (the size of the matrix) and the continuum limit is obtained by going from finitely to infinitely many degrees of freedom. Eq.\eqref{birgf2} has a particularly interesting interpretation when rewritten as
\begin{equation}
g = g_\ast+\#(N^\prime)^{-4/5}\,,
\label{birgf3}
\end{equation}
where $\#$ stands for a constant. This equation has the same structure of eq.\eqref{frg12} if $N^\prime$ is taken as the renormalization group parameter. The critical value $g_{\ast}$ represents thus a fixed point in this abstract renormalization group flow on the matrix size and eq.\eqref{birgf3} is just the linearized flow around such a fixed point. This serves as an inspiration to set up a coarse-graining procedure on the size of the matrix. The Wilsonian perspective would then be realized by integrating out rows and columns of the matrices step by step as discussed in \cite{Brezin:1992yc} and in \cite{Eichhorn:2013isa} in the context of the FRG. In summary, one could discover the double-scaling limit as a fixed-point in the renormalization group flow. Such a coarse-graining procedure does not rely on the introduction of a background which sets up a reference scale. Inhere, the perspective is not the usual one where we flow from ``fast modes" (or high momentum modes) to ``slow modes" (low momentum modes) but rather, we flow from many degrees of freedom to fewer degrees of freedom. In this sense, this is a background-independent coarse-graining procedure. For an extensive discussion on that, we refer to \cite{Eichhorn:2018phj}. 

Having understood how to sum over geometries (and topologies) in two dimensions and how to take a universersal continuum limit, the natural step forward is the generalization to higher dimensions. However, in this case, more limited tools are available. One possibility to perform an accurate calculation is to evaluate the discretized partition function employing Monte-Carlo numerical simulations. Such a program has evolved to the so-called (Causal) Dynamical Triangulations ((C)DT) approach to quantum gravity, see \cite{Ambjorn:2012jv,Ambjorn:2011cg,Ambjorn:2012ij,Ambjorn:2017tnl,Laiho:2016nlp}. Several strong indications point towards a continuum limit which features an extended, four-dimensional geometry at large distances. More refined numerical calculations are the source to establish more robust indications towards the desirable continuum limit. On the other hand, one could try to generalize the matrix model to higher dimensions. This leads to the so-called tensor models, \cite{Ambjorn:1990ge,Godfrey:1990dt,Gross:1991hx,Rivasseau:2011hm,Gurau:2011xp,Gurau:2016cjo,Bonzom:2016dwy}. The Feynman diagrams for such models are dual to discrete geometries and the Feynman expansion corresponds to a sum over geometries and topologies. As before, the fundamental building blocks of geometry correspond to the interacting vertex of random tensors of size $N^\prime$. For a rank-$d$ tensor, each index is associated to a $(d-2)$-subsimplices of a $(d-1)$-simplex, i.e., to the edges of a triangle for rank-3 tensor model. The interaction in this case corresponds to the gluing of such triangles along their edges. In contrast to matrix models, the Feynman expansion of tensor models could not be organized in a $1/N^\prime$ expansion in its original formulation. Nevertheless, a class of tensor models introduced in \cite{Gurau:2010ba,Gurau:2011aq,Gurau:2009tw,Gurau:2011xq,Bonzom:2012hw,Carrozza:2015adg} with a particular type of interactions feature a $1/N^\prime$-expansion leading to the possibility of probing the continuum limit for such models analytically at least for leading order in $1/N^\prime$.

The particular class of tensor models we mentioned are the so-called (un)colored models \cite{Gurau:2011xp}. In these models, the interactions are subject to a $O(N^\prime)^{\otimes d}$ $(U(N^\prime)^{\otimes d})$ symmetry for real (complex) tensors. This means that under the transformation 
\begin{equation}
T_{a_1\ldots a_d}\,\,\,\to\,\,\, T^{\prime}_{a_{1}\ldots a_d} = \sum_{b_1 \ldots b_{d}}O^{(1)}_{a_{1}b_{1}}\ldots O^{(d)}_{a_{d}b_{d}}T_{b_1 \ldots b_d}\,,
\label{birgf4}
\end{equation}
the tensor model action is left invariant, where $O^{(n)}_{ab}$ are orthogonal matrices (the case of complex tensors is completely analogous). Such a symmetry restricts how tensors $T_{a_1\ldots a_d}$ (which do not have any symmetry under index permutation) should be contracted: the first index of a tensor should contract with the first index of another tensor, the second index should contract with the second index and so on. 

In complete analogy to matrix models, it is possible to set up a coarse-graining procedure based on the size of the tensor. The Wilsonian renormalization is realized by integrating out ``layers" of the tensors step by step and suitable universal continuum limits can be discovered by finding fixed points along the renormalization group flow. This is a subject which has witnessed growing interest in the last years not only for pure tensor models \cite{Eichhorn:2014xaa,Eichhorn:2018ylk,Eichhorn:2018phj}, but also, group field theories, see \cite{Benedetti:2014qsa,Benedetti:2015yaa,Geloun:2015qfa,Geloun:2016qyb,Carrozza:2016vsq,Lahoche:2016xiq,Carrozza:2016tih,Carrozza:2017vkz,BenGeloun:2018ekd,Lahoche:2018vun,Lahoche:2018oeo,Lahoche:2018ggd,Lahoche:2018hou,Lahoche:2019vzy,Krajewski:2015clk,Krajewski:2016svb}. See also \cite{Eichhorn:2017bwe,Eichhorn:2019xav} for discussions on the application if coarse-graining techniques to causal sets, another discrete approach to quantum gravity.

As discussed in \cite{Eichhorn:2014xaa,Eichhorn:2018ylk,Eichhorn:2018phj}, we can adapt the FRG to tensor models. In fact, all the derivation of the flow equation outlined Sect.~\ref{FRGoverview} is valid for tensor models provided that the parameter $s$ is identified with the parameter which measures the size of the tensor $N$. Explicitly, the flow equation for tensor models is expressed as
\begin{equation}
\partial_t \Gamma_N [T] = N\partial_N \Gamma_N [T] = \frac{1}{2}\mathrm{Tr}\left[\left(\frac{\partial^2\Gamma_{N}[T]}{\partial T_{a_1\ldots a_d}\partial T_{b_1\ldots b_d}}+R_N\delta_{a_1 b_1}\ldots \delta_{a_d b_d}\right)^{-1}\partial_t R_N\right]\,,
\label{birgf5}
\end{equation}
where $R_N$ is the regulator kernel. Hence, beta functions for the tensor model couplings can be derived. Since the theory is background independent and the coarse-graining parameter $N$ is dimensionless, the assignment of canonical dimensions to the couplings is performed by requiring that the system of beta functions is autonomous. This can be achieved by demanding that the canonical dimensions are such that the beta functions can be expanded in powers of $1/N$. For sufficiently large $N$, the leading order terms form an autonomous system.

In \cite{Eichhorn:2018ylk} a systematic search of fixed points was performed for a rank-3 real tensor model. Different classes of fixed points were found: one class which features dimension reduction, i.e., at the fixed point, the tensor model can be reduced to a matrix model and, therefore, cannot be associated to three-dimensional quantum gravity and other class which does not reduce to a matrix model. The computation of the critical exponents which characterize the continuum limit associated with the second class of fixed point led to two relevant directions with values of order one and a third relevant direction with value close to 0.4. These were obtained in a relatively simple truncation for the effective average action meaning that such numbers come with systematic errors and a check of apparent convergence of the results is definitely needed. Nevertheless, these numbers are compatible with those obtained for the critical exponents associated to the fixed point obtained in the asymptotic safety program in three dimensions \cite{Biemans:2016rvp}. It is definitely too early to state that such fixed points belong to the same universality class and that tensor models and asymptotic safety could be completely different point of views of the same physics. Nevertheless, this opens an exciting possibility to explicitly compare critical exponents in more refined calculations and check whether this premature conclusion could be indeed realized. Results obtained in the simplest truncation for rank-4 models were also reported in \cite{Eichhorn:2018phj}.

\section{Visions: bridging the gap between different approaches to quantum gravity} \label{Visions}

The asymptotic safety scenario for quantum gravity as well as tensor models were discussed as different perspectives to make sense of the path integral for quantum gravity. They intend to perform the path integral using different mathematical tools. In the first case, the ``sum over geometries" is performed by a gauge-fixed sum over metrics. This is achieved, in practice, by introducing an auxiliary background metric and summing over the quantum fluctuations (not necessarily small) around it. On the other hand, tensor models provide a lattice-like discretization of geometries. In both cases, a continuum limit which allows for the removal of a cutoff (in the case of asymptotic safety, a UV cutoff in momentum space and in tensor models, the lattice spacing). This is translated to the existence of fixed points in the renormalization group flow. We have discussed how the FRG can be a versatile tool, adapted to both cases, to discovery such fixed points. It allows for the calculation of critical exponents which characterize the universality class of the underlying continuum limit. 

As pointed out, there are preliminary indications that those theories could belong to the same universality class. If this is really true, then they correspond to different mathematical formulations of the same physical theory, i.e., they lead to the same observables. Although it is very challenging to show this is really the case, the application of the FRG to these theories is relatively simple and more refined results could be obtained in the near future establishing a closer relation between them or revealing they are incompatible. We advocate that such an interplay is crucial for further developments in those fields. As a concrete example, the equivalence between asymptotically safe quantum gravity and tensor models could establish background independence of the Reuter fixed point since in tensor models, no preferred background is employed. Conversely, probing phenomenological aspects such as quantum-gravitational effects in black holes \cite{Pawlowski:2018swz,Adeifeoba:2018ydh,Platania:2019kyx,Bosma:2019aiu}, cosmology \cite{Bonanno:2017gji,Gubitosi:2018gsl} and matter is much easier in the usual QFT for metrics setting. We should exploit the advantages of each formulation to answer different questions. 

Renormalization group techniques are extremely useful and important in many different areas of physics. This article intended to briefly comment on some aspects of the renormalization group in quantum gravity. It could work as a unifying tool which brings together different approaches to quantum gravity giving a coherent and consistent picture of quantum spacetime. Our vision is that such cross-fertilization between different approaches mediated by the renormalization group will bring key new insights to our understading of the quantum microstructure of spacetime.

\section*{Acknowledgments}
I would like to thank Astrid Eichhorn for many inspiring discussions on the topic and the organizers of the conference ``Progress and Visions in Quantum Theory in View of Gravity: Bridging foundations of physics and mathematics" in Leipzig (2018) for the invitation. This work was supported by the DFG through the grant Ei/1037-1.

\bibliography{refs}

\end{document}